\documentclass[aps,prb,twocolumn,10pt]{revtex4}

\usepackage{graphicx}
\usepackage{amsmath}
\usepackage{amssymb}
\usepackage{amsfonts}
\usepackage{bm}
\usepackage{psfrag}
\usepackage{pstricks}

\usepackage{color}

\begin{document}

\title{Structural, electronic, and optical properties of $m$-plane InGaN/GaN quantum wells: Insights from experiment and atomistic theory}

\author{S. Schulz,$^{1}$ D.~P. Tanner,$^{1,2}$ and E.~P. O'Reilly$^{1,2}$}
\affiliation{$^{1}$Photonics Theory Group, Tyndall National
Institute,
Dyke Parade, Cork, Ireland\\
$^{2}$Department of Physics, University College Cork, Cork, Ireland}

\author{M.~A. Caro$^{3,4}$}
\affiliation{$^{3}$Department of Electrical Engineering and
Automation, Aalto University, Espoo 02150, Finland\\$^{4}$COMP
Centre of Excellence in Computational Nanoscience, Aalto University,
Espoo 02150, Finland}

\author{T. L. Martin}
\affiliation{Department of Materials,
            University of Oxford,
            Oxford OX1 3PH, UK}
\author{P. A. J. Bagot}
\affiliation{Department of Materials,
            University of Oxford,
            Oxford OX1 3PH, UK}
\author{M. P. Moody}
\affiliation{Department of Materials,
            University of Oxford,
            Oxford OX1 3PH, UK}
\author{F. Tang}
\affiliation{Department of Materials Science \& Metallurgy,
            University of Cambridge,
            Cambridge, CB3 0FS, UK}
\author{J. T. Griffiths}
\affiliation{Department of Materials Science \& Metallurgy,
            University of Cambridge,
            Cambridge, CB3 0FS, UK}
\author{F. Oehler}
\affiliation{Department of Materials Science \& Metallurgy,
            University of Cambridge,
            Cambridge, CB3 0FS, UK}
\author{M. J. Kappers}
\affiliation{Department of Materials Science \& Metallurgy,
            University of Cambridge,
            Cambridge, CB3 0FS, UK}
\author{R. A. Oliver}
\affiliation{Department of Materials Science \& Metallurgy,
            University of Cambridge,
            Cambridge, CB3 0FS, UK}
\author{C. J. Humphreys}
\affiliation{Department of Materials Science \& Metallurgy,
            University of Cambridge,
            Cambridge, CB3 0FS, UK}
\author{D. Sutherland}
\affiliation{School of Physics and Astronomy,
            Photon Science Institute, Alan Turing Building,
            University of Manchester,
            Manchester, M13 9PL, UK} 
\author{M. J. Davies}
\affiliation{School of Physics and Astronomy,
            Photon Science Institute, Alan Turing Building,
            University of Manchester,
            Manchester, M13 9PL, UK} 
\author{P. Dawson}
\affiliation{School of Physics and Astronomy,
            Photon Science Institute, Alan Turing Building,
            University of Manchester,
            Manchester, M13 9PL, UK}

\begin{abstract}
In this paper we present a detailed analysis of the structural,
electronic, and optical properties of an $m$-plane (In,Ga)N/GaN
quantum well structure grown by metal organic vapor phase epitaxy.
The sample has been structurally characterized by x-ray diffraction,
scanning transmission electron microscopy, and 3D atom probe
tomography. The optical properties of the sample have been studied
by photoluminescence (PL), time-resolved PL spectroscopy, and
polarized PL excitation spectroscopy. The PL spectrum consisted of a
very broad PL line with a high degree of optical linear
polarization. To understand the optical properties we have performed
atomistic tight-binding calculations, and based on our initial atom
probe tomography data, the model includes the effects of strain and
built-in field variations arising from random alloy fluctuations.
Furthermore, we included Coulomb effects in the calculations. Our
microscopic theoretical description reveals strong hole wave
function localization effects due to random alloy fluctuations,
resulting in strong variations in ground state energies and
consequently the corresponding transition energies. This is
consistent with the experimentally observed broad PL peak.
Furthermore, when including Coulomb contributions in the
calculations we find strong exciton localization effects which
explain the form of the PL decay transients. Additionally, the
theoretical results confirm the experimentally observed high degree
of optical linear polarization. Overall, the theoretical data are in
very good agreement with the experimental findings, highlighting the
strong impact of the microscopic alloy structure on the
optoelectronic properties of these systems.
\end{abstract}

\date{\today}


\pacs{78.67.De, 73.22.Dj, 73.21.Fg, 77.65.Ly, 73.20.Fz, 71.35.-y}

\maketitle

\section{Introduction}

(In,Ga)N/GaN quantum well (QW) structures grown on polar $c$-plane
substrates are the building blocks for nitride-based light emitters
operating in the blue spectral region.~\cite{NaSe93,NaMu94} However,
when trying to push the emission wavelength into the green to yellow
spectral region the device performance is hampered, among other
factors, by the presence of strong electrostatic built-in fields,
arising in part from the strain-dependent piezoelectric polarization
and in part from the strain-independent spontaneous
polarization.~\cite{BeFi97} One consequence of the built-in field is
the so-called quantum confined Stark effect (QCSE) which results in
a shift of the emission to longer wavelength.~\cite{ImKo98,LeGr98}
Also the intrinsic field leads to a spatial separation of electron
and hole wave functions, causing a reduced radiative recombination
rate,~\cite{ImKo98,FiBe99} an effect that can be particularly
undesirable for high-efficiency optoelectronic devices. To
circumvent these effects arising from the intrinsic built-in fields,
which fundamentally are caused by the growth along the polar $c$
axis, significant research has been directed towards the fabrication
of semi- and nonpolar
structures.~\cite{WaBr2000,KiSc2007,ScKn2007,KaHo2007,GuBo2008,Paskova2008,BaDa2009,ScBa2010,ZhTa2011,SeBa2011,PaTa2012,Scho2012,MaBr2012,KuSc2014,SuZh2015}
In the case of semi- and nonpolar planes the $c$ axis is at a
nonvanishing angle with respect to the growth direction. In
semipolar planes residual built-in fields are expected and observed,
even though these fields are significantly
reduced.~\cite{ZhTa2011,Scho2012} In terms of built-in field
reduction, nonpolar QW systems are ideal since in these structures
the $c$-axis lies in the growth plane. Thus, in a perfect nonpolar
system, there is no discontinuity in the built-in polarization
vector field at the heterointerfaces, leading ideally to a
field-free system.~\cite{WaBr2000,ScKn2007} Consequently, in such a
structure the radiative recombination rate should be much higher
than in polar QW structures. Additionally, (In,Ga)N QWs grown on
nonpolar planes offer the possibility of acting as highly efficient
sources of linearly polarized
light,~\cite{GuBo2008,Paskova2008,MaBr2012} which may be of
practical use in, e.g., back-lit liquid crystal
displays.~\cite{Paskova2008} The potentially high degree of optical
linear polarization (DOLP) in a nonpolar QW originates mainly from
differences in the effective hole masses along the growth direction,
which leads to a lifting of the degeneracy of the highest lying
$p$-like valence bands.~\cite{SuBr2003,MaBr2012}

However, the detailed consequences of random alloy fluctuations on
the ideally zero built-in field and the optoelectronic properties of
nonpolar systems have been largely ignored. Experimental studies on
$c$-plane (In,Ga)N/GaN QW systems reveal that random alloy
fluctuations lead to carrier localization effects, which can
dominate the electronic and optical properties of these
structures.~\cite{WaJi2012,HaWa2012} Clear experimental indications
of the importance of alloy fluctuations on the properties of
nonpolar (In,Ga)N QWs have also been
presented.~\cite{SuBr2003,SeBa2011,SuZh2015} To fully understand the
structural, electronic, and optical properties of nonpolar (In,Ga)N
QWs requires the use of a range of advanced experimental and
theoretical techniques. For example, 3D atom probe tomography (APT)
can be used to help to identify whether the QW should be described
by a clustered or random alloy.~\cite{GaOl2007,TaZh2015} This
structural information can then form the basis of detailed
theoretical modeling. However, from a theoretical viewpoint, polar
and nonpolar (In,Ga)N QWs are widely treated as homogeneous systems
described by average material parameters. Obviously, these
approaches completely neglect wave function localization arising
from effects that could be attributed to the microscopic alloy
structure. Recently, continuum-based approaches have been modified
to include random alloy
fluctuations.~\cite{FuKa2008,WaGo2011,YaSh2014} Even though such an
approach captures some of the localization effects introduced by
alloy fluctuations it cannot reveal the microscopic origin of
localization effects, including in particular the presence of
In-N-In-N chains as shown by density functional theory
(DFT).~\cite{LiLu2010,ScMa2015SPIE} Unfortunately, realistic QW
systems cannot be treated in the framework of DFT due to the very
large number of atoms ($> 10^4$) required in the calculations.
Additionally, especially for nonpolar systems where the strong
\emph{macroscopic} electrostatic built-in field is absent, DFT
calculations would have to be coupled with solving the
Bethe-Salpeter equation or time-dependent DFT would have to be used
to account for excitonic effects. These effects could be very
important when comparing the results with photoluminescence (PL)
measurements.~\cite{Best2009} To be able to take all these factors
into account, namely the large number of atoms, the atomistic
description of the electronic structure, plus Coulomb correlation
effects, semiempirical approaches, such as
tight-binding~\cite{MoNi2010,DeMa2015,ScCa2015} (TB) or empirical
pseudopotential methods,~\cite{Best2009,MoGa2012} coupled with
configuration interaction (CI)
schemes,~\cite{FrFu99,ScSc2006,Best2009} are required.

In this work we combine advanced experimental and theoretical
methods to analyze the structural, electronic and optical properties
of nonpolar $m$-plane (In,Ga)N/GaN QWs. Our approach allows us to
clarify the impact of alloy fluctuations on the properties of
nonpolar (In,Ga)N structures. We have conducted a comparative study
of the predicted and measured optical properties of a 2 nm thick
$m$-plane (In,Ga)N/GaN QW structure grown on a freestanding GaN
substrate. The heterostructure was grown by metal organic vapor
phase epitaxy (MOVPE) and characterized by x-ray diffraction (XRD),
scanning transmission electron microscopy (STEM), and 3D APT
measurements. The detailed optical characterization of the sample
was carried out by PL, polarized PL excitation (P-PLE) spectroscopy,
and PL time decay measurements. The measured optical properties are
compared with theoretical results. The theoretical framework is
based on an atomistic TB model coupled with CI calculations to
account for Coulomb effects. Our microscopic model takes input from
experimentally determined structural properties and accounts for
local strain and built-in field fluctuations arising from alloy
fluctuations in the QW region.

The TB model assumes that the In atom distribution inside the QW is
close to a random alloy, and we briefly discuss this assumption in
the context of available APT data. The optical characterization
reveals strong Stokes shifts and broad PL linewidth, which is
indicative of strong carrier localization effects. Additionally, and
in stark contrast to $c$-plane systems, our time-resolved
measurements exhibit decay transients that are single exponential.
Also, we find in general an extremely high DOLP ($>90\%$). Using the
experimentally obtained structural information as input for
atomistic TB calculations we find very good agreement between theory
and the optical spectroscopy measurements.

The theoretical analysis reveals strong hole wave function
localization, which leads to a broad emission spectrum. When we
include the Coulomb interaction between the electrons and holes in
our calculations we find that electron wave functions localize about
the hole wave functions, giving rise to localized excitons. This
finding is consistent with the experimentally observed form of the
decay transients. Also, the calculated PL transition energy is in
good agreement with the experimentally measured value. Additionally,
analyzing the wave function character of the valence band edge (VBE)
shows a high DOLP similar to the experimental data. Overall, this
combined experimental and theoretical analysis provides clear
insights into the basic physical properties of $m$-plane
(In,Ga)N/GaN QWs.

The paper is organized as follows. The details of the growth of the
sample are given in Sec.~\ref{sec:Sample_growth}. This is followed
by the description of the applied structural characterization
techniques, the results obtained, and how they are implemented in
the theoretical framework (Sec.~\ref{sec:Structural_Charac}). In
Secs.~\ref{sec:Optical_Setup} and~\ref{sec:Results_optical} we
describe the setup and the results of the experimental optical
characterization of the $m$-plane (In,Ga)N/GaN QWs, respectively. In
Sec.~\ref{sec:Theory_Frame} the ingredients of our theoretical
framework are discussed, while Sec.~\ref{sec:Results_Theory}
presents the theoretical results and the detailed comparison with
the experimental data. Finally, we summarize our results in
Sec.~\ref{sec:Conclusion}.

\section{Sample Growth details}
\label{sec:Sample_growth}

The $m$-plane (In,Ga)N/GaN multiple-QW structure studied was grown
by MOVPE in a Thomas Swan $6\times2$ in close-coupled showerhead
reactor on a freestanding $m$-plane (1-100) GaN substrate. The
misorientation of the substrate is $2^\circ$ $\pm 0.2^\circ$ in the
negative $c$ direction. The substrate, provided by Ammono S.
A.,~\cite{KuZa2012} has a negligible basal plane stacking fault
density and a threading dislocation density of $<$ 5$\times10^4$
cm$^{-2}$. Onto this freestanding substrate an initially undoped 2
$\mu$m thick homoepitaxial GaN buffer layer was grown. For this
growth step trimethylgallium (TMG) and ammonia (NH$_3$), with H$_2$
as the carrier gas, were used. A reactor pressure of 100 Torr at
1050$^\circ$C was applied. Subsequently five (In,Ga)N QWs were grown
with a quasi-2T method, discussed in detail
elsewhere.~\cite{PePa2007} For the QW growth a reactor pressure of
300 Torr was applied. TMG, trimethylindium (TMI), and NH$_3$ have
been used as precursors, with N$_2$ as the carrier gas. The (In,Ga)N
QWs were grown at a temperature of 735$^\circ$C. After the growth of
each QW period a 1 nm GaN cap was grown at the (In,Ga)N growth
temperature. The flows of TMG and NH$_3$ were kept constant during
the 90 s temperature ramp to 860$^\circ$C. After this 90 s
temperature ramp the growth of the barrier material was completed.

\section{Structural Characterization: Setup, results and input into theoretical framework}
\label{sec:Structural_Charac}

Having presented growth details of the sample, we now turn to the
analysis of its structural properties. In
Sec.~\ref{sec:Structural_Charac_SetUp_Results} we briefly introduce
the experimental techniques applied to analyze the structural
properties of the $m$-plane sample and the outcome of these studies.
We discuss in Sec.~\ref{sec:Structural_Charac_Theory} how the
experimental results serve as input for theoretical supercell (SC)
structures, which underlie the atomistic calculations of the
electronic and optical properties of the sample in question.

\subsection{Structural characterization: Experimental setup and results}
\label{sec:Structural_Charac_SetUp_Results}

To analyze the structural properties of the $m$-plane (In,Ga)N/GaN
QW system, different techniques have been applied to obtain a
comprehensive picture. The $\text{In}_{x}\text{Ga}_{1-x}$N QW and
barrier widths as well as the In fraction ($x$) of the sample were
determined by XRD. The sample was analyzed using a high-resolution
x-ray diffraction MRD diffractometer from Panalytical equipped with
a symmetric 4-bounce monochromator and a triple-axis analyzer to
select the CuK$\alpha$1 wavelength. $\omega-2\theta$ scans of the
brightest symmetric reflection (1-100) were used to analyze the
multiple QWs, with a range of 10 degrees in $\omega$. The measured
QW width was 2.0$\pm$0.3 nm with a barrier width of 6.1$\pm$0.3 nm.
The In composition was determined to be 18.3$\pm$0.7\%. The sample
was also characterized with an FEI Osiris fitted with an
extreme-Schottky field emission gun operated at 200 keV. The
measured QW and barrier widths confirmed the XRD measurements.

\begin{figure}[t!]
\includegraphics{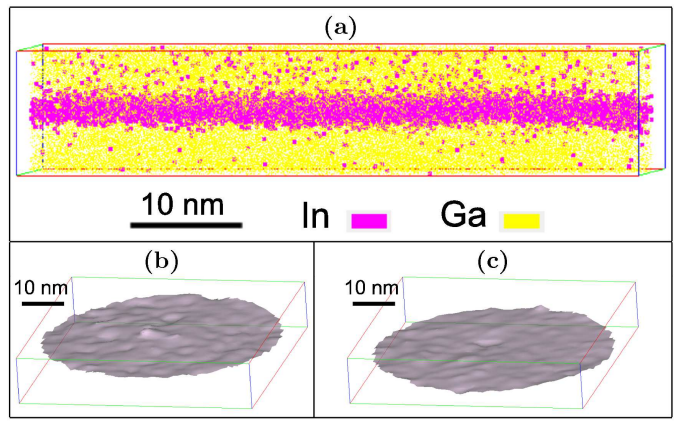}
\caption{APT analysis of the first QW (closest to the substrate);
box size is $55 \times 55 \times 11$ nm$^3$. (a) 3D reconstruction
of the first QW showing 30\% Ga atoms and 50\% In atoms; (b) upper
and (c) lower interfaces of the QW rendered as isosurfaces at 5\% In
content.} \label{fig:APT_results}
\end{figure}

APT experiments were conducted in pulsed laser mode with a pulse
energy of 0.012 nJ using a Cameca LEAP 3000X HR, where the base
temperature of the sample was set at $\sim$30 K and a constant
detection rate of 0.01 atoms per pulse was employed. The IVAS
software package (CAMECA Version 3.6.6) was used to carry out APT
reconstruction and analysis was informed by the thicknesses of the
(In,Ga)N and GaN layers measured by XRD and transmission electron
microscopy (TEM) and also the geometry of the tip as measured by
scanning electron microscopy (SEM). The In content of the sample was
analyzed using concentration profiles through each QW computed using
a proximity histogram (proxigram) approach as described in
Ref.~\onlinecite{TaZh2015}. The mean In fractions across the five
QWs based on the maxima of the In concentration profiles is
17.0$\pm$0.6\%, which is in good agreement with the XRD data. While
frequency distribution analysis of $c$-plane QWs provides no
evidence for the formation of nonrandom indium
clusters,~\cite{GaOl2007} recent studies of nonpolar $a$-plane QWs
with a similar In content (15\%) to the $m$-plane samples studied
here have presented clear evidence for the presence of
clustering.~\cite{TaZh2015}  For the sample under discussion here,
the question of clustering is less clear-cut. Evidence of clustering
is sensitively dependent on the exact parameters of the analysis
used, and the issue is still under investigation.  However, in this
paper we model the QW as a random alloy and assess the extent to
which this model can accurately reproduce the observed optical
properties, which can provide an insight into the relative
importance of slight deviations from randomness in these QWs, if
present. More details of the APT analysis will be reported in
Ref.~\onlinecite{TaZh2015prep}.

An isoconcentration surface analysis of the APT data similar to that
performed by Galtrey \emph{et al.}~\cite{GaOl2007} has been
performed to assess whether the roughness of the upper interface of
the QWs [at which GaN is grown on (In,Ga)N] is rougher than the
lower interface of the QWs [at which (In,Ga)N is grown on GaN] and
typical data are shown in Fig.~\ref{fig:APT_results} for an indium
concentration at the isosurface of 5\%.  This yields a roughness for
the lower surface of 0.24 nm (root mean square), but a larger
roughness for the upper surface of 0.48 nm, with the increased
roughness largely relating to raised island-like features a few nm
in diameter and 1-2 monolayers in height, consistent with the
monolayer and bilayer well-width fluctuations previously observed on
the $c$ plane. High-resolution scanning transmission electron
microscopy imaging of the surfaces of the QWs also indicates the
existence of surface roughness. Although the projection problem
makes it difficult to distinguish the exact nature of the roughness
observed in STEM, the data are consistent with the APT observations.

\subsection{Structural characterization: Input into the theoretical framework}
\label{sec:Structural_Charac_Theory}

Based on the experimental data discussed in
Sec.~\ref{sec:Structural_Charac_SetUp_Results}, we take the
$m$-plane QW width to be approximately 2 nm. We include in our
calculations disklike 2 monolayer thick well-width fluctuations at
one of the QW interfaces. This is a reasonable representation of the
well-width fluctuations that have been observed in APT and is
consistent with earlier approaches to the modeling of such
features.~\cite{WaGo2011} However, for the $m$-plane QW structure
considered here, we do not expect that 2 monolayer thick well-width
fluctuations are of central importance for the description of the
electronic and optical properties. This assumption is based on the
fact that in an $m$-plane QW the macroscopic field is absent.
Therefore, in contrast to a $c$-plane system, the wave functions in
an $m$-plane system are not localized near the interface between QW
and barrier material, since there is no field causing this form of
localization. Consequently, the wave functions are expected to be
far less sensitive to the shape and diameter of the well-width
fluctuations. As we will show below, this assumption is consistent
with the result of our calculations. The In content in the sample
has been measured to be 18.3$\pm$0.7\% (XRD) and 17.0$\pm$0.6\%
(APT), respectively. For our calculations we have set the QW In
content to 17\%, close to the measured values. We use here SCs with
a size of $10\times9\times10$ nm$^3$, thus containing approximately
82,000 atoms, with periodic boundary conditions. Based on the APT
data discussed in Sec.~\ref{sec:Structural_Charac_SetUp_Results}, we
treat (In,Ga)N as a random alloy. To be able to perform a detailed
comparison with experiment and to allow for reasonable statistical
averages, our calculations have been repeated 75 times for the fixed
In fraction of 17\%. This allows us to realize a large number of
microscopically different random configurations in a large SC, where
we have nominally the same In content in the SC but the positions of
the In atoms have been changed randomly.

\section{Electronic and optical properties: Experiment and theory}
\label{sec:Elec_Optical_Charac}

Having investigated the structural properties of the sample, we turn
now to study the electronic and optical properties of the $m$-plane
(In,Ga)N/GaN QW system. We start in Sec.~\ref{sec:Optical_Setup} by
describing the experimental methods used for the optical
characterization, followed by the discussion of the experimental
results in Sec.~\ref{sec:Results_optical}. We then briefly introduce
the theoretical framework in Sec.~\ref{sec:Theory_Frame}, while
Sec.~\ref{sec:Results_Theory} focuses on the comparison of
theoretical and experimental data.

\subsection{Experimental setup}
\label{sec:Optical_Setup}

The PL and P-PLE studies were carried out either using excitation
from a cw He/Cd laser or using a combination of a 300 W xenon lamp
and monochromator as a fixed or tunable wavelength excitation
source. The sample was mounted in the cryostat so that the $c$ axis
of the GaN was horizontal. The PL from the sample was analyzed by a
0.85 m double-grating spectrometer and a Peltier-cooled GaAs
photomultiplier using standard lock-in detection techniques. The
spectral dependence of the DOLP of the emission was determined by
measuring the PL spectra polarized in the plane parallel
($I_\parallel$) and perpendicular ($I_\perp$) to the $c$ axis of the
sample. Once the spectra are obtained, the DOLP is calculated
from~\cite{SuBr2003}
\begin{equation}
\text{DOLP}=\frac{I_\perp-I_\parallel}{I_\perp+I_\parallel}\,\, ,
\label{eq:DOLP}
\end{equation}
where $I_\perp$ ($I_\parallel$) is the PL intensity for the electric
field $\mathbf{E}$ perpendicular $\mathbf{E}\perp c$ (parallel
$\mathbf{E}\parallel c$) to the $c$ axis.

A time-correlated single-photon counting system was used for the
time decay measurements. For these studies the exciting light was
generated by a mode-locked Ti:sapphire laser with a
frequency-tripled emission output of 4.88 eV. A spectrometer
followed by a micro channel plate detector was used to detect the
emitted light.

\subsection{Experimental results: Optical characterization}
\label{sec:Results_optical}

\begin{figure}[t]
\includegraphics[width=\columnwidth]{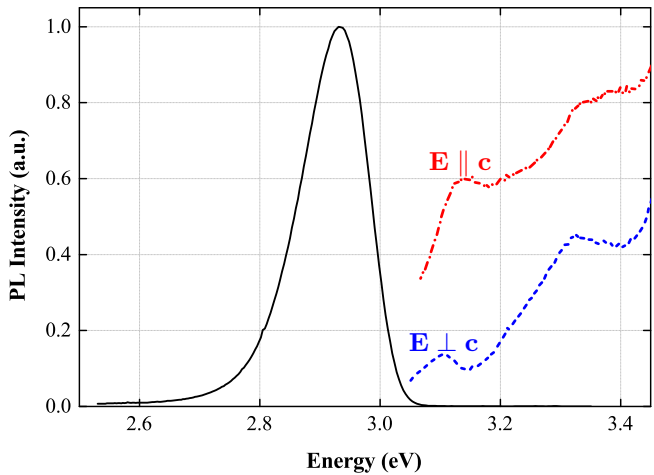}
\caption{Low-temperature (10 K) PL and PLE spectra for the $m$-plane
(In,Ga)N/GaN QW structure under consideration. The PLE spectra are
shown with the plane of polarization of the incident light
perpendicular (blue dashed) or parallel (red dashed-dotted) to the
$c$ axis of the crystal. The PL spectrum is given by the black solid
line. For PL and PLE measurements the excitation spot was 0.02
cm$^2$ with an excitation power density of $\sim$ 10 $\mu$W/cm$^2$.}
\label{fig:PL_exp}
\end{figure}

In this section we describe the results of our PL, P-PLE, and time
decay measurements at 10 K on the $m$-plane (In,Ga)N/GaN QW sample
described above. Figure~\ref{fig:PL_exp} shows the P-PLE spectra
with the plane of polarization of the incident light either
perpendicular ($\mathbf{E}\perp c$) or parallel
($\mathbf{E}\parallel c$) to the $c$ axis of the crystal.
Figure~\ref{fig:PL_exp} also displays the (unpolarized) PL spectrum
of the sample (solid black line). We note several specific aspects
of the PL spectrum. First, the full width at half maximum (FWHM) of
the PL spectrum is 135 meV. This value is much larger than typical
FWHM values found in $c$-plane systems.~\cite{Wats2011} Although it
should be noted that the PL spectrum exhibits an extended low-energy
tail, we have shown previously~\cite{SuZh2015} that in our samples
this low-energy tail arises from recombination in QWs present on
semipolar facets which form at step bunches associated with the
miscut of the GaN substrate. These semipolar QWs have higher QW
width and In content than the nonpolar QWs on the adjoining
$m$-plane terraces on either side of the step, and thus emit at
longer wavelength.  We do not aim to include such structures in our
model or reproduce this aspect of the spectra here. The emission
from the semi-polar QWs is of secondary importance in determining
the magnitude of the spectral width. Second, the energy difference
(Stokes shift) between the peak of the lowest exciton transition in
the P-PLE spectrum and the peak of the PL spectra is 180 meV. We
also note an energy splitting between the two lowest exciton
transitions for the two polarizations of the excitation light of 35
meV. We attribute this splitting mainly to the crystal field
splitting energy in the (In,Ga)N system.~\cite{YaRi2011}

\begin{figure}
\includegraphics[width=\columnwidth]{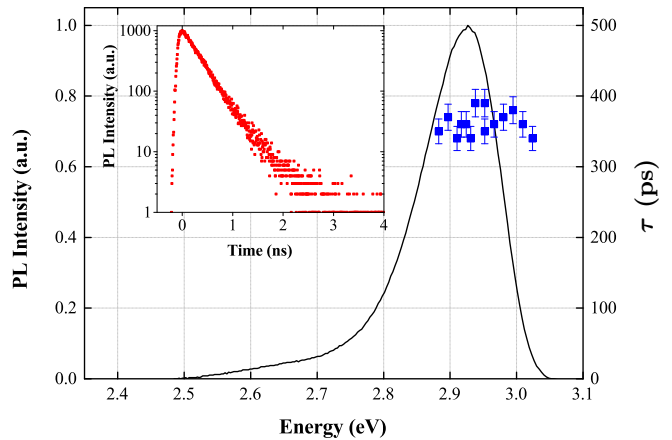}
\caption{(Color online) The unpolarized PL spectrum is displayed by
the solid black line. The data are taken under time-averaged
conditions, where the average power density of the laser source was
1 W/cm$^2$ using 100 fs pulses at a repetition rate of 800 kHz and a
spot size of $2\times10^{-5}$ cm$^2$. The inset shows the
time-dependent PL results. The detection energy for the PL
transients was 2.934 eV. The spectral dependence of corresponding
time constants $\tau$ is given (blue squares). All measurements are
performed at low temperatures (10 K).} \label{fig:TDPL}
\end{figure}

The values of the linewidth and Stokes shift, based on the
assumption that the emission is an intrinsic process, are strongly
suggestive of emission involving strongly localized carriers.  As to
the nature of the localization we anticipate that, similarly to
$c$-plane systems, alloy fluctuations will play a major role.

To help further understand the nature of the recombination we
performed PL time decay measurements across the spectrum. The
results of these measurements are depicted in Fig.~\ref{fig:TDPL}.
Interestingly, our measurements reveal that the PL decays are single
exponentials with characteristic time constants $\tau=350\pm 20$ ps
over the majority of the spectrum. Throughout this work we assume
that the recombination is purely radiative and the values of $\tau$
are the radiative decay constants. This assumption is supported by
temperature-dependent measurements (not shown) and the fact that the
values of $\tau$ found here are similar to those of Marcinkevicius
\emph{et al.}~\cite{MaKe2013}, also measured at low temperature. The
data reported by Marcinkevicius \emph{et al.}~\cite{MaKe2013} at low
temperature were shown to be not influenced by nonradiative
recombination. Only on the low-energy side of the spectra did we
find more slowly decaying emission as we have reported
elsewhere.~\cite{SuZh2015} The emission on the low-energy side is
attributed to semipolar QWs at step bunches described above. The
decay time from the emission associated with these semi-polar QWs
occurs on a longer time scale due to the locally present built-in
field, which separates the electron/hole wave functions.

It is important to note that the radiative decay dynamics observed
here are in stark contrast to $c$-plane (In,Ga)N QW
structures.~\cite{PoBe2000,MoLe2003,GrSo2005,DaDa2014} First, in
polar (In,Ga)N QWs the decay occurs over a much longer time scale
due to the polarization field perpendicular to the plane of the
QWs.~\cite{PoBe2000,GrSo2005,BrPe2010,DaDa2014} Second, the decay
curves are nonexponential due to the variable in-plane separation of
the separately localized electrons and
holes.~\cite{MoLe2003,BrPe2010} Single-exponential decays in
nonpolar QWs have also been reported by other
groups.~\cite{GaSh2009,MaKe2013} The explanation proposed for the
nature of the decay transients is that recombination involves
localized excitons.~\cite{GaSh2009,MaKe2013}

\begin{figure}
\includegraphics[width=\columnwidth]{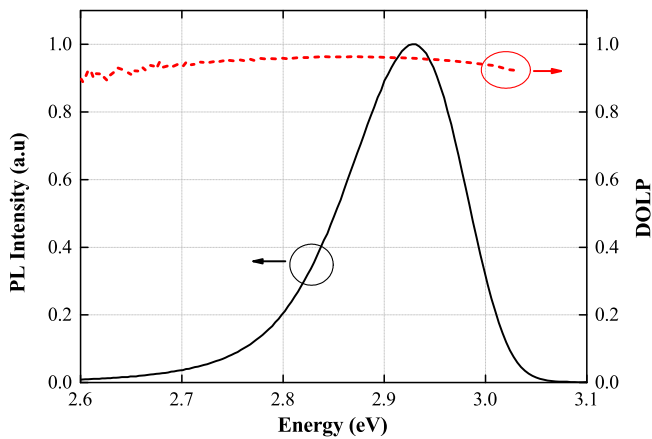}
\caption{(Color online) The (unpolarized) PL emission spectrum for
the $m$-plane (In,Ga)N/GaN QW sample is shown by the black solid
line while the spectral dependence of the DOLP is given by the red
dashed line. All measurements have been performed at low
temperatures (10 K). The cw excitation power density was 12 W/cm$^2$
with an excitation spot of $5\times10^{-5}$ cm$^2$.}
\label{fig:DOLP_exp}
\end{figure}

As discussed in the introduction, nonpolar QWs should exhibit a high
DOLP. This should be particularly the case at 10 K, bearing in mind
the splitting of the lowest valence band states as revealed by the
PLE measurements shown in Fig.~\ref{fig:PL_exp}. The measured DOLP
of the $m$-plane (In,Ga)N/GaN QW sample at 10 K, based on the method
described in Sec.~\ref{sec:Optical_Setup}, is shown in
Fig.~\ref{fig:DOLP_exp} together with the (unpolarized) PL spectrum.
We find a very high DOLP ($> 90$\%), which is in good agreement with
other experimental data on $m$-plane (In,Ga)N/GaN
QWs.~\cite{SuBr2003,KuOk2008,MaBr2012} Theoretically, based on
continuum-like descriptions, at the VBE one would expect a very high
DOLP for several reasons. First, assuming growth along the $y$
direction, the QW confinement lifts the degeneracy of the
$|X\rangle$- and $|Y\rangle$-like valence band states due to their
different effective masses along the growth
direction.~\cite{SuBr2003,ScBa2010,MaBr2012} Second, $|X\rangle$-
and $|Z\rangle$-like states are energetically separated by the
crystal field splitting energy in (In,Ga)N alloys.~\cite{YaRi2011}
Thus, the VBE is expected to be predominately $|X\rangle$-like.
Figure~\ref{fig:DOLP_exp} shows that the DOLP is almost constant
over the entire spectrum. Only on the low-energy side of the
spectrum do we find a small reduction in the DOLP. This reduction is
to be expected as the low-energy part of the PL spectrum originates
from recombination involving semipolar QW states. The detailed
strain state of such QWs on nanoscale facets associated with step
bunches is not straightforward to predict and might cause increased
band mixing effects and consequently a reduced DOLP. A detailed
investigation of these semipolar contributions is beyond the scope
of the present study. Overall, given that the PL linewidth exceeds
the valence band splitting, it is not immediately obvious why the
DOLP, shown in Fig.~\ref{fig:DOLP_exp}, is virtually constant across
the spectrum. One possible answer is that the PL emission spectrum
originates almost entirely from transitions involving localized
states in different potential fluctuations. Therefore, calculations
including random alloy fluctuations are required to shed more light
on this behavior. In the next section we introduce the theoretical
framework used to address these questions.

\subsection{Theoretical Framework}
\label{sec:Theory_Frame}

We use an atomistic TB model to study the electronic and optical
properties of $m$-plane (In,Ga)N-based QWs. This approach allows us
to include (random) alloy fluctuations in the QW region and the
resulting local strain and built-in potential fluctuations on a
microscopic level. The strain field calculations are based on a
modified valence force field model accounting for electrostatic
corrections in addition to bond bending, bond stretching and related
cross-terms.~\cite{ScCa2015} Local built-in field fluctuations are
treated on the basis of our recently developed local polarization
theory.~\cite{CaSc2013local} Strain and built-in field fluctuations
are included in the $sp^3$ TB model presented in
Ref.~\onlinecite{CaSc2013local}. The bulk TB parameters were
determined from fitting TB band structures to hybrid functional
(HSE) DFT band structures, showing a very good agreement between the
TB and HSE-DFT band structures. The TB parameters and the TB model
are discussed in detail in Ref.~\onlinecite{CaSc2013local}. The
application of the TB model to QW structures, including the coupling
of the TB model to CI calculations to include excitonic effects, is
discussed in detail in Ref.~\onlinecite{ScCa2015}. The model has
already been benchmarked against experimental and DFT data on bulk
(In,Ga)N alloys, revealing a very good agreement between experiment,
DFT, and TB results.~\cite{CaSc2013local} Moreover, our recent
calculations on electronic and optical properties of $c$-plane
(In,Ga)N/GaN QWs show also a very good agreement with available
experimental data.~\cite{ScCa2015} The model can directly be
applied, without modification of the theoretical framework, to
$m$-plane structures. Here, only the orientation of the QW structure
in the simulation SC has to be changed compared to a $c$-plane
system. The numerical and structural details of the system
considered here (QW width, In content, number of atoms in the SC, SC
size, etc.) have already been discussed in
Sec.~\ref{sec:Structural_Charac_Theory}.

\begin{figure}[t!]
\includegraphics[width=\columnwidth]{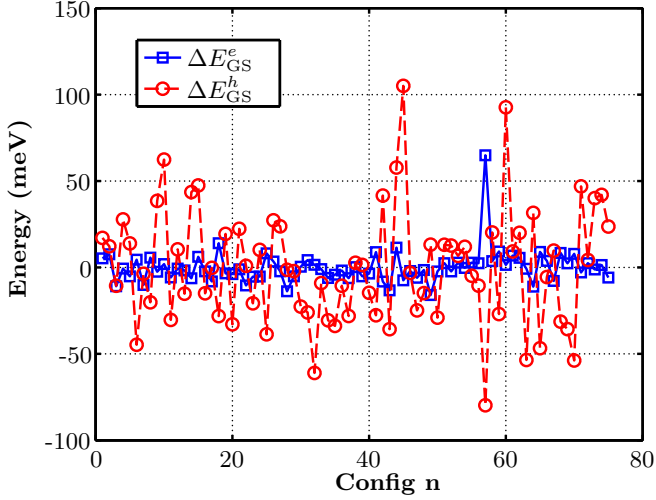}
\caption{(Color online) Variation of the electron $\Delta
E^{e}_\text{GS}$ (blue square) and hole $\Delta E^{h}_\text{GS}$
(red circle) ground state energies with respect to their average
ground state energies $\bar{E}^{e,h}_\text{GS}$ $\left(\Delta
E^{e,h}_\text{GS}=E^{e,h}_\text{GS}(\text{Config}\,\,
n)-\bar{E}^{e,h}_\text{GS}\right)$. The results are shown as a
function of the configuration (Config) number $n$.}
\label{fig:energy_vary}
\end{figure}

\subsection{Theoretical results: Electronic structure and optical properties}
\label{sec:Results_Theory}

Equipped with the knowledge about the experimental data, we present
here the results of the TB calculations.
Figure~\ref{fig:energy_vary} shows the variation of the
single-particle electron $\Delta E^{e}_\text{GS}$ and hole $\Delta
E^{h}_\text{GS}$ ground state energies with respect to the average
ground state energy  for electrons $\bar{E}^{e}_\text{GS}$ and holes
$\bar{E}^{h}_\text{GS}$, respectively, as a function of the
configuration (Config) number $n$ $\left(\Delta
E^{e,h}_\text{GS}=E^{e,h}_\text{GS}(\text{Config}\,\,
n)-\bar{E}^{e,h}_\text{GS}\right)$. From Fig.~\ref{fig:energy_vary}
we can infer that variations in the electron ground state energies
are much smaller than the variations in the hole ground state
energies $\left(\Delta E^{e}_\text{GS}\ll\Delta
E^{h}_\text{GS}\right)$. This is also reflected in the calculated
standard deviations $\sigma$. For electrons we find a standard
deviation of $\sigma^{e}=9.8$ meV, while the standard deviation for
the hole ground state energy is $\sigma^{h}= 33.7$ meV. We have
observed a similar behavior in $c$-plane (In,Ga)N/GaN
QWs.~\cite{ScCa2015}

\begin{figure}[h!]
\includegraphics{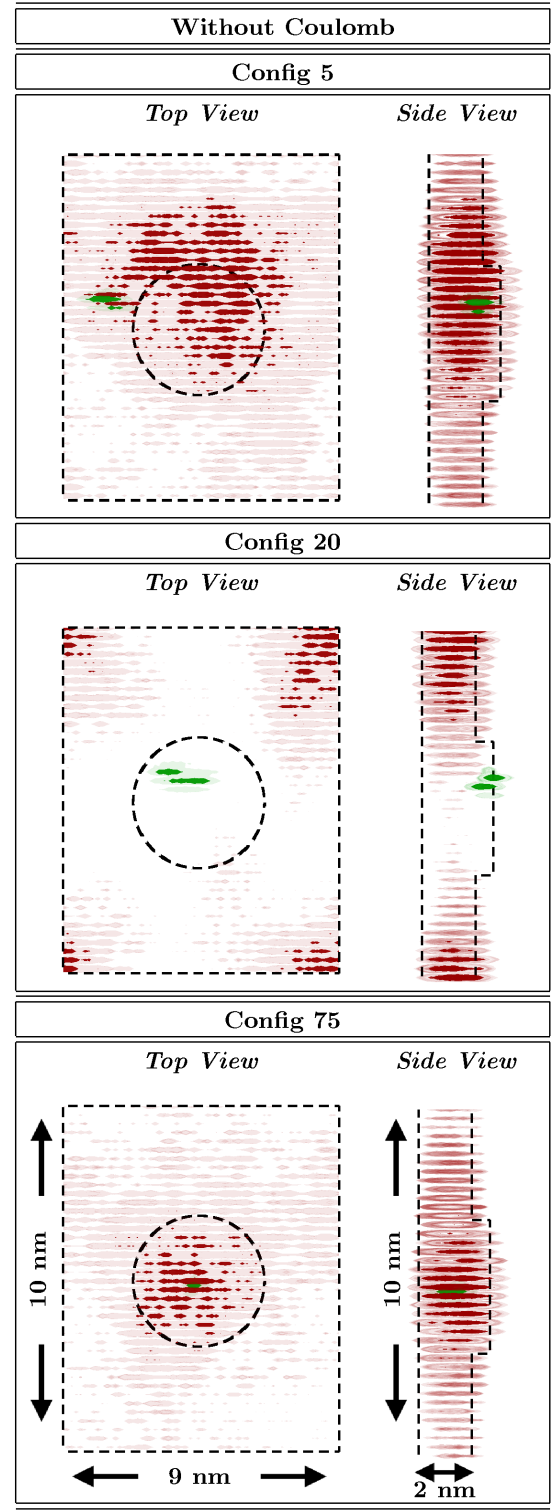}
\caption{(Color online) Single-particle ground state electron (red)
and hole (green) charge densities for configurations 5 (Config 5),
20 (Config 20), and 75 (Config 75). Results are shown for different
view points. Light (dark) isosurfaces correspond to 5\% (25\%) of
the maximum charge density value. Dashed lines indicate the QW
interfaces.} \label{fig:WF}
\end{figure}

The strong variations in the hole ground state energies in $c$-plane
systems are attributed to hole wave function localization effects
due to random alloy fluctuations.~\cite{WaGo2011,YaSh2014,ScCa2015}
To confirm this behavior in a nonpolar system, Fig.~\ref{fig:WF}
shows isosurfaces of the single-particle electron (red) and hole
(green) ground state charge densities for configurations 5, 20, and
75, which reflect typical situations observed in the random alloy as
we will discuss below. The light and dark isosurfaces correspond to
5\% and 25\% of the maximum charge density values, respectively.
From Fig.~\ref{fig:WF} one can infer that the random alloy
fluctuations lead to strong hole wave function localization effects,
independently of the configuration. In general we find that the
alloy fluctuations have a much weaker effect on the electron wave
functions, when compared with the holes. As expected from our
discussion in Sec.~\ref{sec:Structural_Charac_Theory}, 2 monolayer
thick well-width fluctuations do not contribute significantly to
wave function localization effects. Unless very deep well-width
fluctuations are present, which would generate large $c$-oriented
surface areas, we do not expect that well-width fluctuations change
the localization characteristics of electron and hole wave functions
in comparison to the here presented data. Therefore, we conclude
that in $m$-plane QW structures, grown on freestanding GaN
substrates, localization effects can be mainly attributed to alloy
fluctuations. Looking at the displayed configurations in more
detail, for configuration 5, the electron wave function is almost
spread out over the entire QW area, with perturbations introduced by
the alloy fluctuations. We observe slightly stronger perturbations
in the case of configurations 20 and 75 (cf. Fig.~\ref{fig:WF}).
Overall there is a clear difference in the wave function
localization characteristics between electrons and holes. All this
shows that a continuum-based description might give, to a first
approximation, a reasonable description of the ground state electron
wave functions but would fail for the description of the hole ground
states. One should also note that the differences in the
localization features of the different electron single-particle
states, as discussed above, are important. For instance, in
configuration 75 the electron charge density is mainly localized in
the region where the hole ground state is localized. Consequently,
one could expect a higher wave function overlap between electron and
hole. For configuration 5, the electron ground state charge density
has stronger contributions (dark red isosurfaces) in areas spatially
separated from the strongly localized hole ground state. This is
even more pronounced for configuration 20 (cf. Fig.~\ref{fig:WF}).

\begin{figure}[t]
\includegraphics{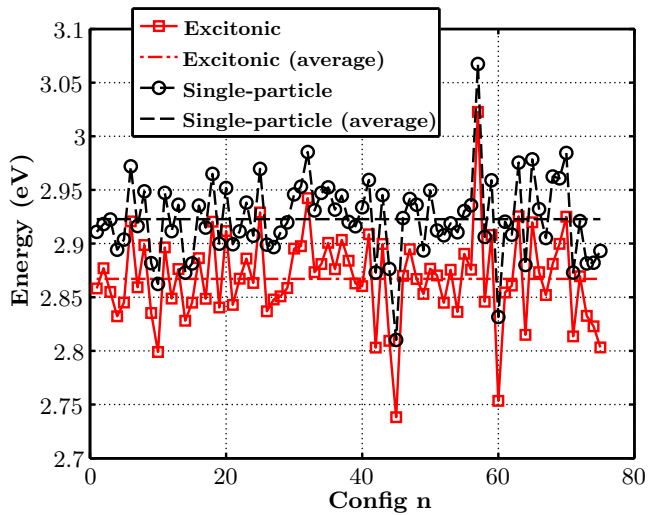}
\caption{(Color online) Single-particle (black circles) and
excitonic (red squares) transition energies as a function of the
configuration number $n$. The average single-particle transition
energies are indicated by the (black) dashed line, while the (red)
dashed-dotted line indicates the average excitonic transition
energy.} \label{fig:Transition}
\end{figure}

In general, the observed asymmetry in the localization
characteristics of electron and hole ground state wave functions is
consistent with DFT-based calculations for \emph{bulk} (In,Ga)N
alloys,~\cite{LiLu2010,ScMa2015SPIE} and explains the observed
asymmetry in the electron and hole ground state energy variations
displayed in Fig.~\ref{fig:energy_vary}. While the local atomic
arrangement (In-N-In chains) is of secondary importance for the
electrons, it is of central importance for the hole ground states,
as shown by DFT-based calculations.~\cite{LiLu2010,ScMa2015SPIE}
Thus, different microscopic arrangements of In atoms will give very
different hole ground state energies. This picture of localized hole
and delocalized electrons is not quite compatible with the
experimentally observed decay times being constant across the
spectrum, as we would still expect some variation in the electron
and hole in-plane wave function overlap leading to a range of
radiative recombination times, i.e., nonexponential decay curves. It
should however be noted that the effect would be much less marked
than in the case of polar QWs where the electrons and holes are
localized separately. We will discuss this discrepancy later.

\begin{figure}[t!]
\includegraphics{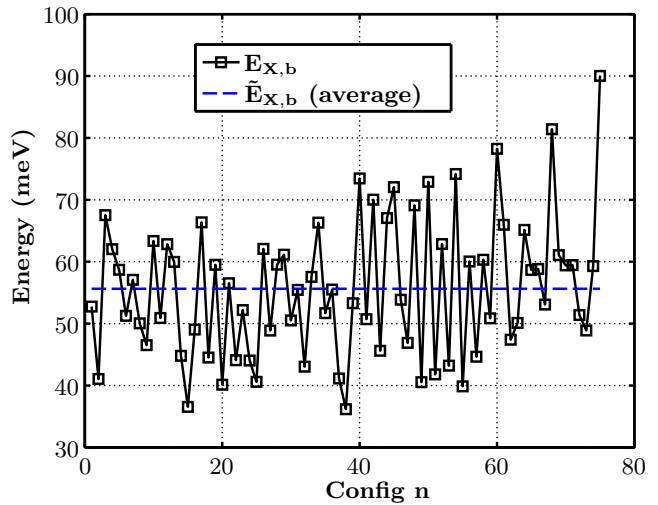}
\caption{Excitonic binding energy $E_{X,b}$ as a function of the
configuration number $n$. The average excitonic binding energy
$\tilde{E}_{X,b}$ is indicated by the dashed line.}
\label{fig:X_binding}
\end{figure}

The variations in the single-particle energies will also affect the
single-particle ground state emission energies. This is shown in
Fig.~\ref{fig:Transition}. The single-particle transition energies
(black circles) show large variations between 2.81 eV and 3.07 eV.
The average single-particle transition energy is around 2.93 eV
(black dashed line). However, to be able to compare the theoretical
ground state emission energies with the experimental PL emission
energies, we have to include Coulomb (excitonic) effects. We use the
CI scheme described in detail in our previous work, to include
excitonic effects in the description.~\cite{ScCa2015} We assume here
a single electron-hole pair. Thus, electron-electron and hole-hole
Coulomb interactions are not considered. We neglect electron-hole
exchange contributions since these are small corrections on the
energy scale relevant for the discussion of our results. To describe
the excitonic many-body wave function we include 5 electron and 15
hole states in the CI expansion. The calculated excitonic ground
state energies (red squares) are shown in Fig.~\ref{fig:Transition}
again as a function of the configuration number $n$. The average
excitonic transition energy is given by the (red) dashed dotted
line. From Fig.~\ref{fig:Transition} we conclude that Coulomb
effects introduce a significant shift and broadening of the spectrum
when compared with the single-particle results. However, when
calculating the excitonic binding energy $E_{X,b}$ as the difference
between the single-particle ground state transition energy and the
excitonic ground state transition energy, we find also large
variations in $E_{X,b}$. The excitonic binding energy $E_{X,b}$ is
displayed in Fig.~\ref{fig:X_binding} as a function of the
configuration number $n$. The calculated values scatter between 36
meV and 90 meV. On average we find here an excitonic binding energy
of approximately 56 meV (dashed line). The difference in the
observed excitonic binding energies can be understood from the
single-particle wave functions shown in Fig.~\ref{fig:WF}. We find
that for configuration 20 the electron wave function, in comparison
to configurations 5 and 75, shows very little charge density in the
spatial region where the hole is localized, leading in the
single-particle picture to a reduced wave function overlap. From
this one could expect a lower exciton binding energy in comparison
to a configuration where electron and hole wave functions are
localized in the same spatial region. As shown in Fig.~\ref{fig:WF},
this is the situation for configuration 75, which has consequently a
very large excitonic binding energy (cf. Fig.~\ref{fig:X_binding}).
The intermediate situation is realized for configuration 5 giving
also an intermediate excitonic binding energy. In general it is
important to note that excitonic binding energies in nonpolar QWs
are much larger than in $c$-plane systems. This effect is expected
due to the absence of the \emph{macroscopic} built-in field in
nonpolar nitride-based QWs.~\cite{WeJi2012,FuKa2008}

\begin{figure}[h!]
\includegraphics{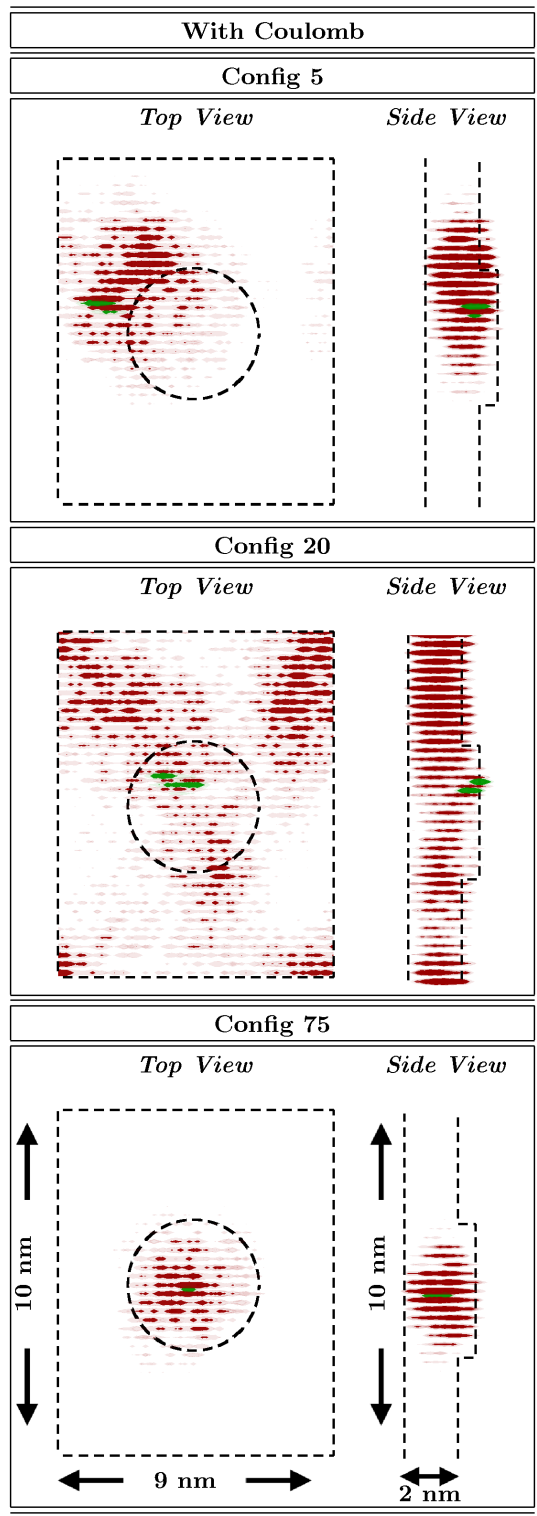}
\caption{(Color online) Ground state electron (red) and hole (green)
charge densities with Coulomb effects included for configurations 5
(Config 5), 20 (Config 20), and 75 (Config 75). Results are shown
for different view points. Light (dark) isosurfaces correspond to
5\% (25\%) of the maximum charge density value. Dashed lines
indicate the QW interfaces.} \label{fig:X_WF}
\end{figure}

To shed even more light onto the impact of the Coulomb interaction
on the wave functions, we use reduced electron and hole density
matrices to visualize the electron and hole densities under the
influence of the Coulomb interaction. In general, the excitonic
many-body wave function $|\psi^{X}\rangle$ can be written as a
\emph{linear combination} of electron-hole basis states:
\begin{equation}
|\psi^{X}\rangle=\sum\limits_{i,j}c^{X}_{ij}\hat{e}^{\dagger}_{i}\hat{h}^{\dagger}_j|0\rangle\,
.
\end{equation}
Here $|0\rangle$ is the vacuum state, $c^{X}_{ij}$ the expansion
coefficient, and $\hat{e}^{\dagger}_{i}$ ($\hat{h}^{\dagger}_{j}$)
denotes the electron (hole) creation operator. Electron and hole
states are denoted by $i$ and $j$, respectively. We can then define
reduced density matrices for electrons and holes. For instance, for
the electrons the density operator $\hat{\rho}^{e}$ is given by:
\begin{equation}
\hat{\rho}^{e}=\sum_{i,i'}|i\rangle\sum_{j}c^{X}_{ij}c^{X*}_{i'j}\langle
i'|=\sum_{i,i'}|i\rangle\rho^{e}_{ii'}\langle i'| \,\, .
\end{equation}
The corresponding electron and hole densities are given by
$\rho^{e}=\langle \mathbf{R}|\hat{\rho}^{e}|\mathbf{R}\rangle$ and
$\rho^{h}=\langle \mathbf{R}|\hat{\rho}^{h}|\mathbf{R}\rangle$,
respectively. Figure~\ref{fig:X_WF} depicts the calculated electron
${\rho}^{e}$ and hole ${\rho}^{h}$ densities for the configurations
5, 20, and 75. We can infer from Fig.~\ref{fig:X_WF} that while the
hole charge density is almost unchanged in comparison to the
single-particle description (cf. Fig.~\ref{fig:WF}), the electron
charge density is significantly affected by the attractive Coulomb
interaction between electron and hole. The electron charge density,
under the influence of the Coulomb interaction, localizes about the
hole for all configurations. Thus our atomistic model predicts
exciton localization in nonpolar (In,Ga)N/GaN QWs. This behavior is
consistent with the experimental observation of single-exponential
PL decay transients discussed in detail in
Sec.~\ref{sec:Results_optical} and observed experimentally by other
groups.~\cite{GaSh2009,MaKe2013} Also the strong wave function
overlap independent of the hole localization is consistent with the
measured decay times being constant over the spectrum.

\begin{figure}[t]
\includegraphics{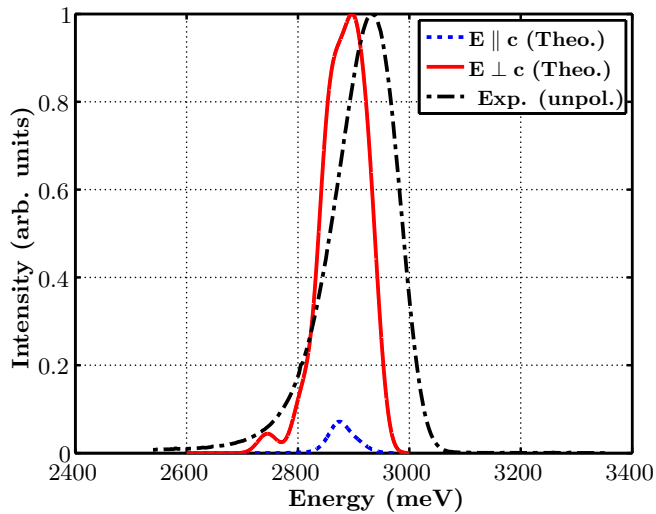}
\caption{(Color online) Calculated excitonic ground state emission
spectrum with light polarization vectors parallel
($\mathbf{E}\parallel c$) and perpendicular ($\mathbf{E}\perp c$) to
the $c$ axis. The (black) dashed-dotted line shows the experimental
(unpolarized) PL spectrum.} \label{fig:PL_Exp_Theo}
\end{figure}

To further compare the theoretical results with the experimental PL
data, we have calculated the excitonic ground state emission
spectrum following Ref.~\onlinecite{ScCa2015}. Since the experiment
indicates a large DOLP ($>90\%$) we have performed calculations for
two different light polarization vectors $\mathbf{e}_{p,i}$, namely
$\mathbf{e}_{p,\perp}=(1,0,0)^T$ and
$\mathbf{e}_{p,\parallel}=(0,0,1)^T$. This means that the selected
electric field $\mathbf{E}$ is perpendicular to the $c$ axis
($\mathbf{e}_{p,\perp}$) and parallel to the $c$ axis
($\mathbf{e}_{p,\parallel}$), respectively. In the calculations we
have assumed growth along the $y$ axis; therefore the chosen light
polarization vectors reflect the experimental setup described in
Sec.~\ref{sec:Optical_Setup}. The calculated excitonic ground state
emission spectrum is shown in Fig.~\ref{fig:PL_Exp_Theo} for
$\mathbf{e}_{p,\perp}$ (red solid line) and
$\mathbf{e}_{p,\parallel}$ (blue dashed line) together with the
experimental (unpolarized) PL emission spectrum (black dashed-dotted
line). Several different features are visible in
Fig.~\ref{fig:PL_Exp_Theo}.

First of all we find a good agreement between the calculated (solid
red) and the experimentally determined emission energy (black
dashed-dotted line), given the slight uncertainties in In content
and well width as discussed in
Sec.~\ref{sec:Structural_Charac_SetUp_Results}. Second, we find also
theoretically a very broad emission spectrum. For $\mathbf{E}\perp
c$ ($\mathbf{e}_{p,\perp}$), the theoretically determined FWHM is
approximately 101 meV. Experimentally we find a FWHM value of 135
meV. Different factors might contribute to the observed differences
between theory and experiment. For example, even though 75 different
microscopic structures may appear a large number, it could be the
case that even more configurations have to be considered to fully
resolve the measured FWHM, bearing in mind the large variations
between different microscopic configurations (cf.
Fig.~\ref{fig:energy_vary}). Additionally, if subtle nonrandom
clustering effects exist, they may contribute to the broadening of
the PL linewidth. However, the theoretically determined value of 101
meV for the FWHM is in reasonable agreement with the experimental
data. Third, Fig.~\ref{fig:PL_Exp_Theo} shows that there is a large
difference in the calculated intensities for $\mathbf{e}_{p,\perp}$
and $\mathbf{e}_{p,\parallel}$. In the theoretical analysis, the
intensities are normalized to the intensity of
$\mathbf{e}_{p,\perp}$. If we calculate the DOLP from our
theoretical spectrum via Eq.~(\ref{eq:DOLP}) based on the maximum
intensities for $\mathbf{e}_{p,\perp}$ and
$\mathbf{e}_{p,\parallel}$, we find a value of approximately 87\%,
which is slightly smaller than the experimental values ($>90$\%).
\begin{figure}
\includegraphics{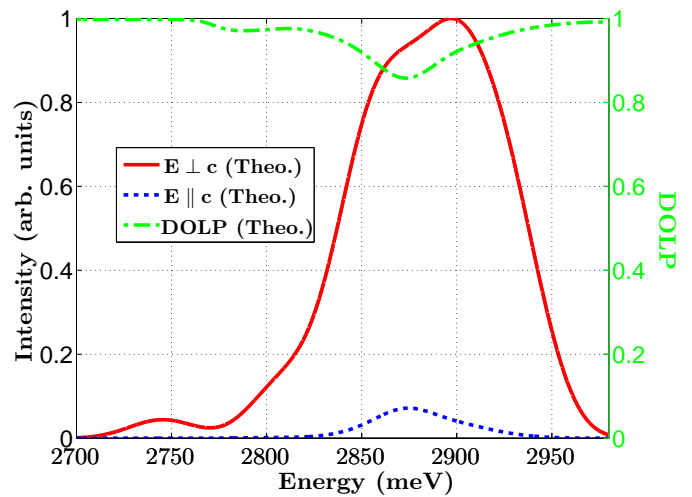}
\caption{(Color online) Calculated excitonic emission spectrum for
the light polarization perpendicular (red solid line) and parallel
(blue dashed line) to the $c$ axis. The spectral dependence of the
corresponding DOLP, calculated according to Eq.~(\ref{eq:DOLP}), is
shown by the (green) dashed-dotted line.} \label{fig:DOLP_Theory}
\end{figure}
To compare our theoretical results in more detail with the
experiment, we have also calculated the spectral dependence of the
DOLP from our theoretical emission spectra for the two light
polarization configurations depicted in Fig.~\ref{fig:PL_Exp_Theo}.
Using Eq.~(\ref{eq:DOLP}), our theoretical results for the spectral
dependence of the DOLP are shown in Fig.~\ref{fig:DOLP_Theory}.
Similarly to Fig.~\ref{fig:DOLP_exp}, the DOLP (green dashed-dotted
line) is shown together with the excitonic emission spectrum for
$\mathbf{E}\perp\mathbf{c}$ (red solid line) and
$\mathbf{E}\parallel\mathbf{c}$ (blue dashed line). When comparing
our theoretical data with the experimental data shown in
Fig.~\ref{fig:DOLP_exp}, the theoretical results show slightly lower
values than the experiment, plus the calculated DOLP is not as
constant as the experimental data across the spectrum. Again, even
though 75 configurations may appear a large number, we show below
that some of the structure in the DOLP spectrum of
Fig.~\ref{fig:DOLP_Theory} is due to a small number of exceptional
states; more configurations would be required to reliably treat the
importance of such states. This is beyond the scope of the present
study, since the present analysis gives already, in general, a good
description of the experimentally observed spectral dependence of
the DOLP.

\begin{figure}
\includegraphics{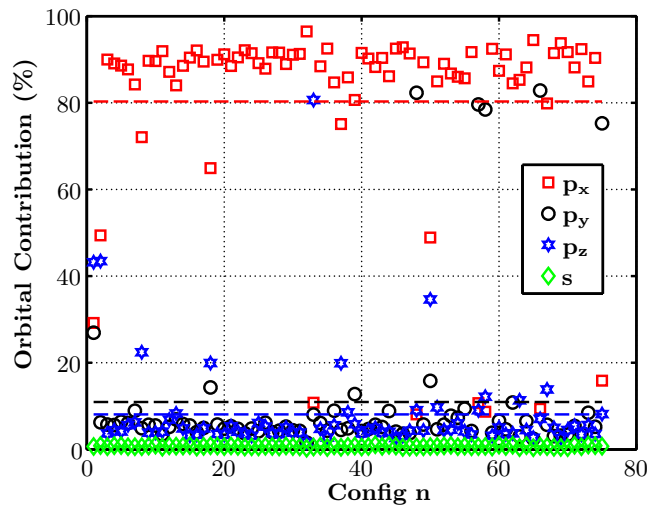}
\caption{(Color online) Orbital contributions to the VBE as a
function of the $n$ different microscopic configurations considered
here.} \label{fig:Orbital_Contrib}
\end{figure}

All in all the theoretical calculations indicate that random alloy
fluctuations, explicitly taken into account in our model, affect the
DOLP only slightly. The origin of the calculated high DOLP can be
further understood by looking at the orbital character of the hole
ground state/VBE state. The outcome of such an analysis is displayed
in Fig.~\ref{fig:Orbital_Contrib}, where the orbital contributions
to the VBE are shown as a function of the configuration number $n$.
From a continuum-based calculation, neglecting the weak spin-orbit
coupling, one would expect, due to the differences in the valence
band effective masses along the growth direction and the positive
crystal field splitting energy in GaN and InN, that the VBE is
dominated by a single-orbital type ($p_x$- or $p_y$-like orbitals).
Obviously such an analysis neglects the effects of alloy
fluctuations. However, we can infer from
Fig.~\ref{fig:Orbital_Contrib} that the VBE state in the different
microscopic configurations is mainly made up of contributions from a
single orbital type, in this case from $p_x$-like orbitals. Thus,
for the orbital character of the hole ground state the microscopic
configuration is, in general, of secondary importance, although we
note that there is an enhanced $p_z$-like character and a very low
$p_x$-like character in about 10\%-20\% of the structures studied.
The dominance of the $p_x$ character of the VBE explains the
calculated high DOLP, which is in good agreement with the
experimental data. Furthermore, as we know from
Fig.~\ref{fig:energy_vary} the hole ground state energies vary
significantly between different configurations. This gives rise to
the broad emission spectrum shown in Fig.~\ref{fig:PL_Exp_Theo}.
Since Fig.~\ref{fig:Orbital_Contrib} reveals that the orbital
contribution to the VBE state is for the most part independent of
the configuration number $n$, all these findings in combination
explain why we observe only a weak spectral dependence of the DOLP
displayed in Fig.~\ref{fig:DOLP_Theory}, in line with the
experimental results displayed in Fig.~\ref{fig:DOLP_exp}.

\section{Conclusion}
\label{sec:Conclusion}

In summary, we have presented a detailed experimental and
theoretical analysis of the structural, electronic, and optical
properties of $m$-plane (In,Ga)N/GaN QWs. The structural
characterization using XRD, STEM, and APT provide us with consistent
parameters for the QW composition and dimensions and suggest that we
may reasonably model the (In,Ga)N as a random alloy, although we are
not able to rule out the possibility of some subtle deviations of
the In distribution from randomness. In addition to the structural
characterization, we used PL, P-PLE, and time-resolved PL
measurements to analyze the optical properties of the system under
consideration. Our results show a high DOLP, single-exponential PL
decay transients, strong Stokes shifts, and a very broad PL
linewidth. The strong Stokes shift and the broad PL line width are
indicative of strong carrier localization effects, while
single-exponential PL decay transients in nonpolar (In,Ga)N QWs are
usually attributed to exciton localization effects. To shed further
light on the experimentally observed properties, we have employed an
atomistic TB model to achieve a microscopic description of the
electronic structure of the $m$-plane system including, on an
atomistic level, strain and built-in field variations arising from
the considered random alloy fluctuations. To be able to compare with
the measured optical spectra, our model includes also excitonic
effects via the CI scheme. The results of our calculations reveal
strong hole wave function localization effects originating from
random alloy fluctuations. We find that alloy fluctuations have a
much weaker effect on the electron wave functions. The observed
localization effects lead to a significant broadening of the
single-particle ground state energies and therefore the
corresponding transition energies. When including Coulomb effects in
the calculations we observe compared to $c$-plane systems strongly
increased excitonic binding energies. This can be attributed to the
absence of the macroscopic built-in field in the nonpolar $m$-plane
system studied here. Coulomb effects lead also to the situation
where the electron wave function localizes about the hole. In other
words, the theoretical calculations predict exciton localization
effects, which are consistent with the single-exponential decay of
the PL transients observed in the experiment. Moreover, when
calculating the excitonic ground state emission spectrum we find a
good agreement with the experimental PL spectra both in terms of
FWHM and PL peak position energy. Additionally, when calculating the
DOLP, the theoretical data are in good agreement with the
experimental result of a very high DOLP. Our findings indicate that
the DOLP is, in general, only slightly affected by random alloy
fluctuations. However, since the observed PL line width is larger
than the measured valence band splitting, wave function localization
effects due to random alloy fluctuations are required to explain the
spectral independence of the DOLP, given that the theoretical
analysis shows that these localized states are dominated by a single
orbital type. Overall, this combined experimental and theoretical
analysis provides clear insight into the basic physical properties
of $m$-plane (In,Ga)N/GaN QWs.

\begin{acknowledgments}
This work was supported by Science Foundation Ireland (Projects No.
13/SIRG/2210 and 10/IN.1/I2994), the United Kingdom Engineering and
Physical Sciences Research Council (Grant Agreements No.
EP\textbackslash J001627\textbackslash 1 and EP\textbackslash
J003603\textbackslash 1), the European Union 7th Framework Programme
DEEPEN (Grant Agreement No. 604416), and in part by the European
Research Council under the European Community's 7th Framework
Programme (FP7/2007-2013)/ERC Grant Agreement No. 279361 (MACONS).
We would like to acknowledge the help of D. Haley, University of
Oxford, on the APT results. S. S. acknowledges  computing  resources
provided by Science Foundation Ireland (SFI) to the Tyndall National
Institute and by the SFI and Higher Education Authority Funded Irish
Centre for High End Computing. Supporting data may be accessed via
the following link:
https://www.repository.cam.ac.uk/handle/1810/252770.
\end{acknowledgments}

\bibliographystyle{apsrev}
\bibliography{../../../phdstef}

\end{document}